\documentstyle[12pt]{article}

\newcommand{\bbc}{\begin{center}}
\newcommand{\eec}{\end{center}}

\begin{document}

\vspace {0.3in}

\begin{center}
\Large\bf Investigation of phase-slip-like resistivity in underdoped YBCO \\
\vspace {0.3in} {\large M. M. Abdelhadi and J.A. Jung  } \\
\vspace {0.1in}
 \small\em Department of Physics, \\
 \small\em University of Alberta, Edmonton, Alberta Canada T6G 2J1.
\end{center}
\vspace {0.1in}

\begin{abstract}
We investigated the anomalous peak resistivity below the onset
$T_c$ in underdoped YBCO, reminiscent of that observed in 1D wires
of conventional superconductors. We performed measurements of the
angular dependence of resistivity $\rho(\theta)$ in a magnetic
field and the temperature dependence of resistivity $\rho(T)$,
which exhibit a peak for $\textbf{B} \parallel$ ab-planes. This
peak in $\rho(T)$ disappears for $\textbf{B} \parallel$ c-axis.
The width of the corresponding maximum in $\rho(\theta)$ at
$\theta=0^o$ ($\textbf{B} \parallel$ ab-planes) decreases with
increasing c-axis component of the field ($B \sin \theta$). The
maximum in $\rho(\theta)$ and $\rho(T)$ decreases with an
increasing applied transport current. We analyzed the data using
three different models of resistivity based on 2D resistor array,
flux motion, and thermally activated phase-slips. Numerical
calculations suggest that in  a filamentary underdoped system, the
phase-slip events could produce the anomalous resistivity close to
$T_c$.
\end{abstract}

\bbc (PACS numbers: 74.76.-w, 74.40.+k,74.72.Bk) \eec
\newpage

\section{INTRODUCTION}

Observation of a large resistive peak in the temperature
dependence of resistivity $\rho(T)$ just below the onset $T_c$ has
been reported in crystals of high $T_c$ superconductors (HTSC)
like $(Nd,Pr)_{1.85}Ce_{0.15}CuO_{4-y}$ \cite{ref:Crusellas},
YBCO(123) \cite{ref:Mosqueira}, and BSCCO(2212) \cite{ref:Han}. In
all these cases, the magnitude of the resistive peak is higher
than the resistivity at the onset $T_c$. The peak shows an
anomalous behavior in a magnetic field applied along the c-axis.
Its magnitude decreases with an increasing field, and in high
enough fields the peak is completely suppressed. The applied
transport current has the same effect on the peak i.e the peak's
magnitude decreases with an increasing applied current.

These phenomena are of considerable interest because of their
striking qualitative similarity to those observed in conventional
superconductors (LTSC) like superconducting mesoscopic Al wires
\cite{ref:Santhanam,ref:Moshchalkov}, thin films of Al
\cite{ref:Kwong,ref:Spahn}, $(NbV)N$, NbV, VN, (NbTi)N
\cite{ref:Vaglio}, and disordered metallic glasses of
$Zr_{60}Cu_{60}$ \cite{ref:Nordstrom}.  Mosqueira et al
\cite{ref:Mosqueira} reported the resistive anomalous peak in
$\rho_{ab}(T)$ of YBCO crystals of $T_c(R=0)=89 K$. This anomaly
was eliminated by successive annealing of the sample in oxygen.
This annealing also led to an increase of $T_c(R=0)$ up to 90.3 K
(close to the optimal doping). The authors concluded that the peak
could be related to very small $T_c$-inhomogeneities non-uniformly
distributed in the crystal. They performed computer simulations of
the temperature dependence of the anomalous resistivity using the
model of two-dimensional electric circuit: an array of resistors
whose resistivity depends on temperature. The non-uniformly
distributed $T_c$-inhomogeneities were introduced by assuming that
these resistors have different (higher or lower) $T_c$. The
authors stated that the uniformly distributed
$T_c$-inhomogeneities at large length scales broaden the resistive
transition only, and do not produce a peak. Similar approach was
introduced earlier by Vaglio et al \cite{ref:Vaglio} to explain
the resistance-peak anomaly in non-homogeneous thin films of
(NbV)N, NbN, VN, and (NbTi)N. They concluded that "the current
redistribution" caused by the sample's inhomogeneity, is
responsible for the observed phenomena.

Current redistribution effects in an inhomogeneous sample were
also considered by Nordstrom and Rapp \cite{ref:Nordstrom} in
their interpretation of the resistive peak anomaly in
superconducting amorphous thick films of $Zr_{60}Cu_{40}$. Kwong
et al \cite{ref:Kwong} observed an anomalous peak in the resistive
transition of a 2D 25 nm thick aluminium film containing regions
of different but comparable, transition temperatures. Disordered
regions of lower $T_c$ were produced by the reactive-ion etching
process. Their observation seems to support earlier
interpretations based on $T_c$-inhomogeneities and current
redistribution effects. The authors stated, however, that the
anomaly could originate from a discontinuity of the
superconducting potential at the normal-superconducting metal
(N-S) interface, and for superconducting electrodes placed
sufficiently close to the interface, this potential exceeds the
normal-state value. Spahn and Keck \cite{ref:Spahn} found that the
anomaly appears in 2D Al films with thickness between 13 and 40
nm. They argued that this effect could be caused by an interaction
between the superconducting fluctuations and the conduction
electrons.

Extensive studies of the resistive anomaly were also performed on
1D Al strips with a width less than the coherence length and the
magnetic penetration depth, by Santhanam et al
\cite{ref:Santhanam} and Moshchalkov et al \cite{ref:Moshchalkov}.
Santhanam et al argued that the Al wire could be treated (at
temperatures close to $T_c$) as a coherent region comprising
normal (N) and superconducting (S) phases. The resulting N-S
interface gives rise to a quasiparticle charge imbalance induced
by the bias current, and consequently to the observed changes in
resistivity. Moshchalkov et al performed quantitative analysis of
the anomaly using Langer-Ambegaokar (LA) \cite{ref:Langer} and
McCumber-Halperin (MH) \cite{ref:McCumber} models of the thermally
activated phase-slips of the superconducting order parameter.
LA-MH models were adopted with the modification which assumes that
in quasi-1D superconducting wires the normal current and the
supercurrent can only flow in series, and the total resistance is
the sum of the normal resistance $R_N$ and the phase-slip
resistance $R_S$. Good quantitative agreement between the
experimental data and the calculated resistance R(T) was obtained.

Crusellas  et al \cite{ref:Crusellas} and Han et al \cite{ref:Han}
proposed that the anomaly in $\rho_{ab}(T)$ of
$(Pr,Nd)_{1.85}Ce_{0.15}CuO_{4-y}$ and BSCCO(2212) crystals is the
manifestation of a quasireentrant behavior, which results from the
intrinsic granularity. Han et al rejected the explanation based on
non-uniformly distributed $T_c$-inhomogeneities
(Ref.\cite{ref:Mosqueira}), because of the observation of an
anomalous peak in the I-V characteristics, which were measured at
different magnetic fields. However, Crusellas et al
\cite{ref:Crusellas} stated that the anomaly is strongly
influenced by the distribution of defects, after it was observed
that high temperature annealing reduces the size of the resistive
anomaly in $(Pr,Nd)_{1.85}Ce_{0.15}CuO_{4-y}$ crystals.

Briefly, the interpretation of the anomalous resistive peaks in
HTSC concentrates on two possible sources of this effect:
non-uniformly distributed $T_c$-inhomogeneities and intrinsic
granularity. The explanation of the anomalous resistivity in LTSC
films took into account the effects of $T_c$-inhomogeneities (and
related current redistribution), N-S interfaces, and the
interaction between superconducting fluctuations and the
conduction electrons. It was also suggested that the anomaly in 1D
LTSC (Al) strips (wires) originates from the presence of N-S
interfaces and/or thermally activated phase-slips of the order
parameter.

These various interpretations are the source of a number of
unanswered questions:

\begin{enumerate}
  \item According to Browning et al \cite{ref:Browning} in YBCO single crystals of
  $T_c$=93 K and transition width of $\Delta T_c$=0.2 K, large
  variation in the oxygen content $7-\delta$ can occur across the
  sample as revealed by high resolution scanning x-ray
  diffractometry (which was performed using $10 \mu m$ wide x-ray
  beam). $7-\delta$ in these crystals ranges between 6.80 and
  7.00, which corresponds to a change of $T_c$ by about 10 K. In
  spite of these non-uniform $T_c$-inhomogeneities, the crystals
  have small resistivities ($\rho \simeq 40 \mu \Omega cm$ at 100
  K) and do not show any resistive peak anomalies at the onset
  $T_c$ in $\rho(T)$. These results throw doubt on whether the
  2D resistive model alone (as proposed in Ref.\cite{ref:Mosqueira} for YBCO) can
  explain the observed anomalies.
  \item The resistive  anomalies observed in LTSC films and wires
  (strips) are similar and their interpretation suggest the link between
  the presence of inhomogeneities ($T_c$-inhomogeneities, N-S
  interfaces) and the superconducting fluctuations, including
  phase-slips of the order parameter. Could this explanation be
  also applied to HTSC?
  \item Moshchalkov et al \cite{ref:Moshchalkov}introduced the phase slip resistivity (
  according to the 1D LA-MH model \cite{ref:Langer,ref:McCumber}) combined with the
normal state
  resistivity in order to explain the anomalous resistive peak in
  1D aluminum wires. Experiments by Browning et al [see (1)] suggest
  filamentary phase separation and filamentary flow of the
  current in some YBCO crystals with sharp superconducting
  transitions, which do not show resistive anomalies. Does this
  mean using the analogy to LTSC that the presence of
  non-uniformly distributed inhomogeneities in HTSC is the necessary but
  not sufficient condition to observe the resistive anomaly?
  What is the other condition? Could this be a 1D current flow in
  an inhomogeneous system?
  \item What is the contribution of the magnetic flux motion
  (pinning) to the observed resistive anomalies in HTSC?
\end{enumerate}

In order to answer these questions new experiments are needed. We
decided to perform measurements of the angular dependence of
resistivity in a magnetic field. This decision was stimulated by
the experiments done on $Pr_{1.85}Ce_{0.15}Cu_{4-y}$ crystals
\cite{ref:Crusellas}, in which the resistive anomaly was
investigated for two different directions of the applied magnetic
field, namely along the c-axis and along the ab-planes. The effect
of the magnetic field on the anomaly was completely different for
these two orientations. Taking into account the fact that the
presence of the non-uniform $T_c$-inhomogeneities is the necessary
but not sufficient condition to observe the resistive anomalies,
we investigated the temperature dependence of resistivity on a
large number of YBCO samples. We decided to study YBCO thin films,
both optimally doped and underdoped, because they are readily
accessible from different research groups and can be deposited on
various substrates using several different deposition techniques.
The bridges for the resistive measurements can be made relatively
easily on thin films using standard photolithographic techniques.
The resistive peak anomaly and the reduction of its magnitude with
an increasing magnetic field and an increasing applied current,
were observed in an underdoped film after investigation of fifteen
YBCO films of $T_c$ ranging between 79 and 90.5 K. This film was
then used to perform detailed measurements of the angular
dependence of resistivity in a magnetic field. The resulting
experimental data were analyzed using three different models: two
dimensional resistor model, magnetic flux motion model, and the
LA-MH thermally activated phase -slip theory.

\section{ EXPERIMENTAL PROCEDURE}

\subsection{Sample Preparation}
C-axis oriented YBCO thin films were prepared using off-axis rf
magnetron sputtering and laser ablation from stoichiometric
$YBa_2Cu_3O_{7-\delta}$ targets of 99.999\% purity. Films were
deposited on three different types of substrates: $SrTiO_3$,
$LaAlO_3$, and sapphire (with $CeO_2$ buffer layer).

 We investigated fifteen YBCO thin films (both
 underdoped and close to the optimal doping) of various zero-resistance transition
 temperatures (between 79 and 90.5K) and thicknesses (between
 100 and 600nm). YBCO films were patterned, using conventional
photolithography and wet etching technique, into a form of a
$30-60\mu m$ wide and 6.4 mm long strips with six measurement
probes. Large area silver contacts were deposited on the film by
rf magnetron sputtering in order to minimize Joule heating. Copper
leads were attached to silver contacts using mechanically pressed
indium. The distance between voltage probes was 0.4 mm.

The anomalous resistivity was observed in an underdoped YBCO thin
film. This film (140 nm thick), was deposited on a (1000) oriented
sapphire substrate (with $CeO_{2}$ buffer layer) using laser
ablation technique. The sample exhibits a vanishing zero-field
resistivity at $T_{c}=81.7 K$ and has a room temperature
resistivity $\rho_{300K}=34.2 \mu\Omega cm $
($\frac{\rho_{300K}}{\rho_{100K}}=2.4$).

X-ray diffraction (XRD) data of this film showed the pattern of a
characteristic stoichiometric c-axis oriented YBCO film. The data
did not reveal any impurity phases. The XRD data gave a c-axis
lattice spacing of 11.70 \AA, which corresponds to an oxygen
content of about 6.8 and $T_c(R=0)$ of about 80 K
\cite{ref:Andersen}.

\subsection{Measurement Procedure}

The investigation of the resistive anomalies was based on the
following measurements: (a) the measurement of the temperature
dependence of resistivity $\rho(T)$ between room temperature and
$T_c(R=0)$ in zero magnetic field; (b) the measurement of the
temperature dependence of resistivity $\rho(T)$ between the onset
$T_c$ and $T_c(R=0)$ in an external magnetic field applied either
parallel or perpendicular to the ab-planes; (c) the measurement of
the angular dependence of resistivity $\rho(\theta)$ as a function
of the angle $\theta$ between the ab-planes and the direction of
the fixed applied magnetic field at fixed temperatures between the
onset $T_c$ and $T_c(R=0)$; (d) the measurement of $\rho(\theta)$
as a function of the magnitude of the magnetic field \textbf{B}
and the applied transport current density \textbf{J}. The angular
measurements were performed by rotating a copper sample holder
about its vertical axis in a horizontal magnetic field up to 1
Tesla, using a combination of a step-motor and backlash-free gear
reducer. The angle was accurately monitored by an 8000-line
optical encoder attached to the sample, whose angular resolution
was $0.045^{o}$. The film was mounted with the c-axis
perpendicular to the sample holder's vertical axis of rotation,
which allowed one to change the magnetic field direction in a
plane parallel to the c-axis.

Resistivity was measured using the standard dc four-probe method.
The current was applied to the sample in the form of short pulses
(of duration less than 200 ms) in order to reduce Joule heating. A
dc current reversal was used to eliminate the thermal emf in the
leads. The voltage was
 measured using a  Keithley 2182 Nanovoltmeter in tandem with
 a Keithley 236 Current Source, with the nanovoltmeter working as the triggering
 unit. The nanovoltmeter was operated in a mode (known as Delta mode) which
 allows the measurement and calculation of the voltage from
 two voltage measurements for two opposite directions of the
 current. Temperature was monitored by a carbon-glass resistance
  thermometer and an inductanceless heater
and was controlled  to better than $\pm 10mK$ for each single
angular sweep in a magnetic field. This was achieved by rotating
the sample very slowly in the magnetic field in order to reduce
variations in the emf in the heater which could disturb the
temperature reading. The term "resistivity" is used in this paper
to denote the quantity $E/J$ (where E is the electric field and J
is the transport current density), and it does not imply  an ohmic
response.

All measurements were carried out with the transport current
\textbf{J} parallel to the ab-planes for two different
orientations of the magnetic field with respect to the current.
For the first one, the field was rotated in a plane perpendicular
to the current direction while for the other one the field was
rotated in a plane parallel to the current and the c-axis
directions [see Fig.\ref{fig:fig1}(a)]. All measurements were done
in a field cooling (FC) regime, with the magnetic field applied to
the sample at a temperature above the  onset $T_{c}$, followed by
a slow cooling down to the required temperature of measurement.

\section{EXPERIMENTAL RESULTS}

\subsection{Temperature dependence of resistivity}

\begin{figure}
\begin{center}
\def\picfilename{Fig.1f.eps}
\input epsf
\epsfxsize = 220pt \epsfysize =230pt
 \epsfbox{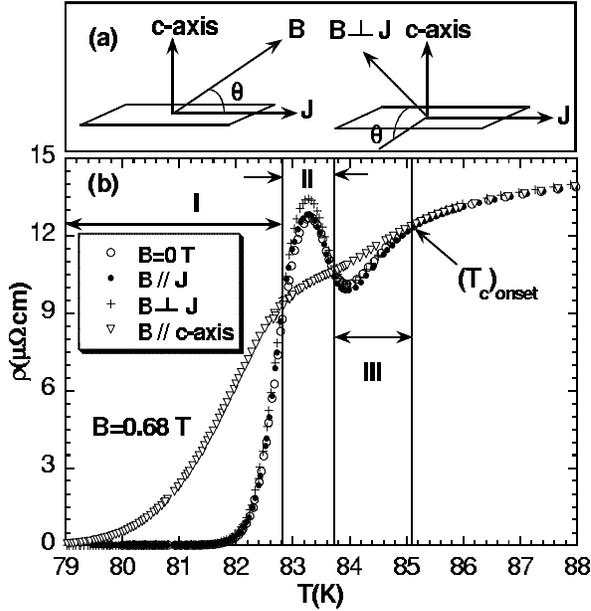}
 \end{center}
\caption{\setlength{\baselineskip}{10pt} (a) Two configurations of
\textbf{B} with respect \textbf{J} that were used during the
measurements of the angular dependence of resistivity
$\rho(\theta)$ in a magnetic field: \textbf{B} is rotated in a
plane parallel to both \textbf{J} and the c-axis (left side), or
\textbf{B} is rotated in a plane parallel to the c-axis but
perpendicular to \textbf{J} (right side). (b) Temperature
dependence of resistivity for YBCO thin film measured in a zero
and 0.68 T  fields at different orientations. Regions I and III
denote the temperature ranges over which $\rho(T)$ is independent
of the magnitude of \textbf{B} and the angle between \textbf{B}
and \textbf{J} for $\textbf{B} \parallel $ ab-planes. In these
regions $\rho(\theta)$ displays a minimum at $\theta=0^{o}$
(\textbf{B} parallel to the ab-planes) [see Fig.\ref{fig:fig2}(a)
and (c)]. Region II represents the temperature range over which
$\rho(T)$ exhibits a peak (of magnitude larger than $\rho(T)$ at
the onset $T_c$), for \textbf{B}=0 and for both $
\textbf{B}\parallel \textbf{J}$ ($\theta =0^o$) and $ \textbf{B}
\perp \textbf{J}$ ($\theta =0^o$) orientations. Rotating the field
from the ab-planes ($\theta =0^o$) towards the c-axis ($\theta
=90^o$), leads to an increase in $\rho(T)$ in regions I and III
 and to a suppression of the peak in region II.}
\label{fig:fig1}
\end{figure}

The temperature dependence of resistivity $\rho$ was measured over
a temperature range of 78-300 K in a zero magnetic field. For a
temperature range (78-90 K) close to $T_c$, $\rho$ was recorded
for different orientations of the magnetic field with respect to
the direction of the current density \textbf{J}.
Fig.\ref{fig:fig1}(a) shows two possible orientations of the
magnetic field \textbf{B} with respect to \textbf{J} and the
ab-plane of the film. The figure on the left illustrates the case
in which the field \textbf{B} was rotated in a plane parallel to
the direction of \textbf{J} while the one on the right represents
the case in which \textbf{B} was rotated in a plane perpendicular
to \textbf{J}. For both orientations \textbf{B} was rotated in a
plane parallel to the c-axis. The first configuration is denoted
as $\textbf{B}\parallel \textbf{J}$ and the second one as $
\textbf{B} \perp \textbf{J}$. Fig.\ref{fig:fig1}(b) shows the
temperature dependence of resistivity $\rho(T)$ for a temperature
range of 79-88 K measured in a zero field and in 0.68 T. The
 measurements of $\rho(T)$ in the field were carried out for the following
 orientations of \textbf{B} with respect to \textbf{J}: $ \textbf{B}
\parallel \textbf{J}$ and $ \textbf{B} \perp \textbf{J}$ with \textbf{B}
parallel to the ab-plane ($\theta =0^o$), and for  $\textbf{B}
\perp \textbf{J}$ with \textbf{B} parallel to the c-axis ($\theta
=90^o$). The onset transition temperature (onset $T_c$) is defined
as the temperature above which the resistivity does not respond to
the change in both magnitude and direction of the magnetic field
[see Fig.\ref{fig:fig1}(b)]. $\rho(T)$ below the onset $T_c$ could
be divided into three regions. Each region is identified according
to the response of $\rho(T)$ to the change in the direction of the
magnetic field from the ab-planes ($\theta=0^o$) to the c-axis
($\theta=90^o$). Region II represents a temperature range between
82.8 K and 83.5 K over which $\rho(T)$ exhibits a peak (of
magnitude larger than that of $\rho(T)$ at the onset $T_c$) in a
zero magnetic field, and for $\textbf{B}
\parallel$ ab-planes with $ \textbf{B}\parallel \textbf{J}$
and $ \textbf{B} \perp \textbf{J}$  orientations. Note a clear
separation between the peak and the onset $T_c$. In this region,
behavior of $\rho(T)$ changes dramatically upon rotating the field
from the ab-planes to the c-axis. In regions I and III, $\rho(T)$
was observed to increase when \textbf{B} is parallel to the
c-axis, while in region II (the peak region) $\rho(T)$ is
completely suppressed by the magnetic field $\textbf{B}
\parallel$ c-axis. For $ \textbf{B}\parallel$ ab-planes, $\rho(T)$
in regions I and III is independent of the magnitude of \textbf{B}
and the angle between \textbf{B} and \textbf{J}. However, in
region II, $\rho(T)$ is independent of the magnitude of \textbf{B}
only for $ \textbf{B}\parallel \textbf{J}$ orientation. In this
region, $ \rho(T)$ was found to decrease with an increasing
applied current density \textbf{J}. The temperature dependence of
resistivity for $\textbf{B}
\parallel c$-axis was measured in different magnetic fields. For
fields above 0.1 T, the peak in region II is completely
suppressed. The resistivity between the onset $T_c$ and the room
temperature exhibits a linear temperature dependence.

\subsection{Angular
dependence of resistivity: Effect of temperature and magnetic
field}

\begin{figure}
\begin{center}
\def\picfilename{Fig.2f.eps}
\input epsf
\epsfxsize = 220pt \epsfysize =300pt \epsfbox{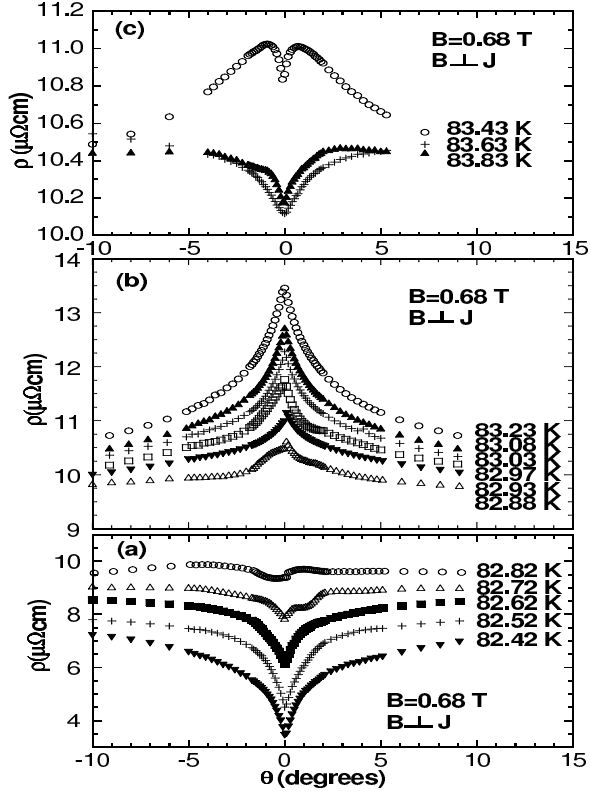}
 \end{center}
\caption{\setlength{\baselineskip}{10pt}$\rho(\theta)$ measured in
0.68 T for a temperature range between 82.52 K and 83.83 K
spanning the three regions I, II, and III. Note the change in
$\rho(\theta)$ at $\theta=0^o$ from a minimum in region I (a) to a
maximum in region II (b) and then back to a minimum in region III
(c). In regions I and III, identical behavior of $\rho(\theta)$
has been observed for $ \textbf{B} \parallel \textbf{J}$ and $
\textbf{B} \perp \textbf{J}$ orientations, whereas in region II
the magnitude of $\rho(\theta)$ depends on those orientations.}
\label{fig:fig2}
\end{figure}

The angular dependence of resistivity $\rho(\theta)$ was measured
in a constant magnetic field at different temperatures in regions
I, II, and III, for both $ \textbf{B}
\parallel \textbf{J}$ and $ \textbf{B} \perp \textbf{J}$ orientations for the
angular range from $-20^{o}$ to $+20^{o}$. The measurements
revealed minima in $\rho(\theta)$ at $\theta=0^o$ in region I and
III, and a maximum  in region II (see Fig.\ref{fig:fig2}).
Fig.\ref{fig:fig2}(a) shows $\rho(\theta)$ in region I as a
function of temperature between 82.42 K and 82.82 K in a magnetic
field of 0.68 T. At a temperature of approximately 82.82 K, which
corresponds to the border line between region I and II in
Fig.\ref{fig:fig1}(b), there is a crossover from a minimum to a
peak in $\rho(\theta)$. This peak grows with an increasing
temperature reaching a maximum value at 83.23 K (see
Fig.\ref{fig:fig2}(b)).

The second crossover from a maximum to minimum can be seen at
83.50 K, which corresponds to the border line between regions II
and III in Fig.\ref{fig:fig1}(b). We have measured $\rho(\theta)$
for both $ \textbf{B} \perp \textbf{J}$ and $ \textbf{B}
\parallel \textbf{J}$ in all three regions. While in regions I
and III the minimum in $\rho(\theta)$ is independent of the
orientation of \textbf{B} with respect to \textbf{J} (i.e for $
\textbf{B} \perp \textbf{J}$ and $ \textbf{B}
\parallel \textbf{J}$), in region II the magnitude of the peak depends on these
orientations and $\rho(\theta)_{B \perp J}>\rho(\theta)_{B
\parallel J}$.

\begin{figure}[h]
\begin{center}
\def\picfilename{Fig.3f.eps}
\input epsf
\epsfxsize = 210pt \epsfysize =150pt \epsfbox{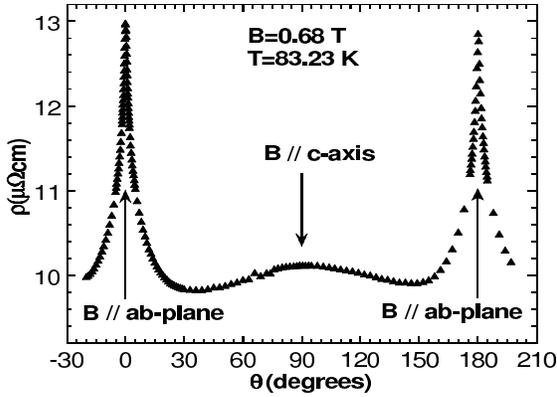}
 \end{center}
\caption{\setlength{\baselineskip}{10pt} $\rho(\theta)$ measured
in 0.68 T at 83.23 K for an angular range $-30^o < \theta <
210^o$. Note the peaks at $\theta=0^o$ and $\theta=180^o$
($\textbf{B} \parallel $ ab-planes) which are approximately 30\%
higher than the maximum at $\theta=90^o$ ($\textbf{B} \parallel $
c-axis).} \label{fig:fig3}
\end{figure}

The measurements of $\rho(\theta)$ over an angular range between
$-30^{o}$ and $210^{o}$ revealed  sharp maxima for $\textbf{B}
\parallel $ ab-planes ($\theta=0^o$ and $\theta=180^o$) and a smaller broad maximum
for $\textbf{B} \parallel $ c-axis ($\theta=90^o$) (see
Fig.\ref{fig:fig3}).  $\rho(\theta)$ at $\theta=0^o$ and
$\theta=180^o$ is about 30\% larger than that for  $\theta=90^o$.
Moreover $\rho(\theta)$ has minima at $\theta =35^{o}$ an
$\theta=145^o$  for all fields.

\begin{figure}[h]
\begin{center}
\def\picfilename{Fig.4.eps}
\input epsf
\epsfxsize = 210pt \epsfysize =150pt \epsfbox{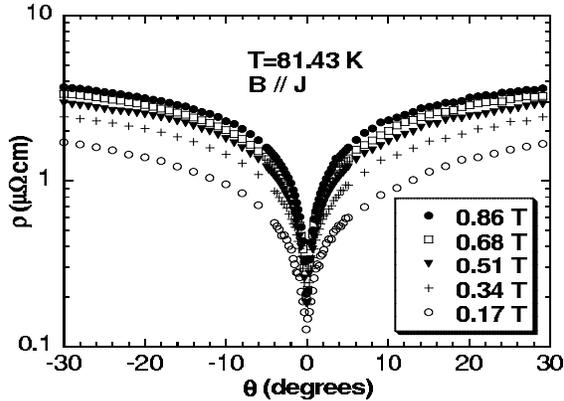}
 \end{center}
\caption{\setlength{\baselineskip}{10pt} (a) Angular dependence of
resistivity measured at a temperature of 81.43 K (region I) in
different fields for  \textbf{B} close to the ab-planes. The data
reveal a minimum in $\rho(\theta)$ at $\theta=0^{o}$ (\textbf{B}
parallel to the ab-planes). The width of the minimum decreases
with an in increasing field. An identical behavior has been
observed for both $\textbf{B} \perp \textbf{J}$ and $\textbf{B}
\parallel \textbf{J}$ orientations.} \label{fig:fig4}
\end{figure}

The angular dependence of resistivity $\rho(\theta)$ was also
measured at a constant temperature in different magnetic field in
regions I, II, and III, for both $ \textbf{B}
\parallel \textbf{J}$ and $ \textbf{B} \perp \textbf{J}$
orientations and for the angular range between $-20^{o}$ and
$+20^{o}$. Fig.\ref{fig:fig4} shows the angular dependence of
resistivity, at a temperature of 81.43 K (region I), measured in
different magnetic fields for both $ \textbf{B}
\parallel \textbf{J}$ and $ \textbf{B} \perp \textbf{J}$
orientations. The data for these two orientations are identical
which implies that $\rho(\theta)$ is independent of the angle
between \textbf{B} and \textbf{J}. $\rho(\theta)$ at $\theta=0^o$
is almost independent of the magnitude of \textbf{B}. The width of
this minimum [defined as half width at half minimum (HWHM)]
decreases from HWHM=$2.3^{o}$ in a field of 0.17 T down to
HWHM=$2.0^{o}$ in 0.86 T. The depth of the minimum increases with
an increasing field. The results of the measurements of
$\rho(\theta)$ in region III is identical in all aspects to those
obtained in region I.

\begin{figure}[h]
\begin{center}
\def\picfilename{Fig.5.eps}
\input epsf
\epsfxsize = 210pt \epsfysize =250pt \epsfbox{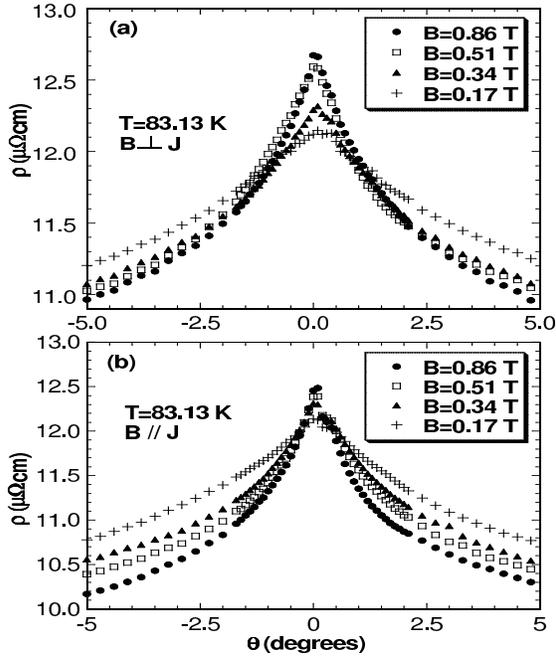}
 \end{center}
\caption{\setlength{\baselineskip}{10pt} Angular dependence of
resistivity $\rho(\theta)$ for $-5^o<\theta <5^o$, measured in
different applied magnetic fields at a fixed temperature of 83.13
K (region II). (a) $\rho(\theta)$ for $\textbf{B} \perp
\textbf{J}$ orientation, where $\rho(\theta)$ increases with an
increasing \textbf{B} for $\theta <1^{o}$ but decreases with an
increasing \textbf{B} for $\theta
>1^{o}$. (b) $\rho(\theta)$ for $\textbf{B}\parallel \textbf{J}$
orientation; $\rho$ at $\theta=0^{o}$ is almost independent of
\textbf{B}. Note that $\rho(\theta)$ decreases with an increasing
\textbf{B} for $\theta > 0.5^{o}$.} \label{fig:fig5}
\end{figure}

Fig.\ref{fig:fig5} presents $\rho(\theta)$ measured at a
temperature of 83.13 K (in region II) in different magnetic fields
for both $\textbf{B} \perp \textbf{J}$ and $ \textbf{B}
\parallel \textbf{J}$ orientations.
The width of the peak in $\rho(\theta)$ decreases with an
increasing \textbf{B} for both $ \textbf{B} \perp \textbf{J}$ and
$ \textbf{B}
\parallel \textbf{J}$ orientations. The magnitude of the peak
in $\rho(\theta)$ for $ \textbf{B}\parallel \textbf{J}$ is almost
independent of the magnitude of \textbf{B}, however it increases
with \textbf{B} for $ \textbf{B} \perp \textbf{J}$. A decrease of
the peak's width with an increasing magnetic field means that
within a certain angular range ($\mid \theta \mid
> 1.5^{o}$ for $\textbf{B} \perp \textbf{J}$ and $\mid \theta \mid > 0.3^{o}$
 for $ \textbf{B} \parallel \textbf{J}$), $\rho(\theta)$ decreases with an
 increasing field.

\subsection{Angular dependence of resistivity: Effect of the
applied current}

\begin{figure}[h]
\begin{center}
\def\picfilename{Fig.6.eps}
\input epsf
\epsfxsize = 210pt \epsfysize =250pt \epsfbox{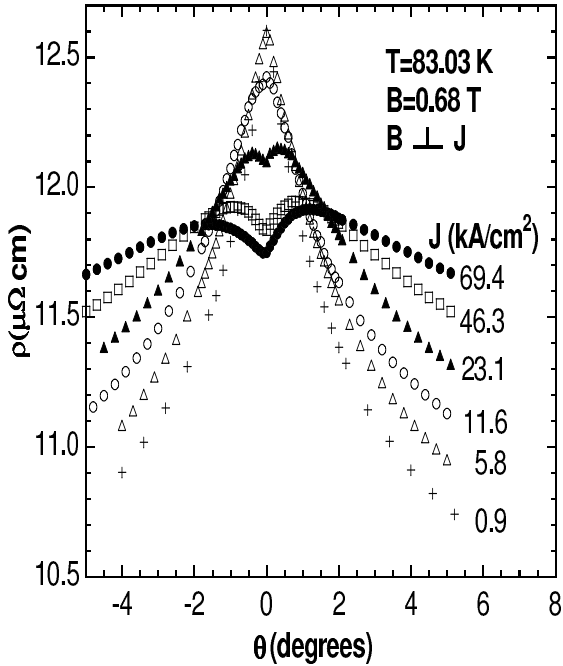}
 \end{center}
\caption{\setlength{\baselineskip}{10pt} Angular dependence of
resistivity $\rho(\theta)$ measured as a function of  applied
current density \textbf{J} for a field of 0.68 T and a temperature
of 83.03 K (region II) . For angles $\mid \theta \mid
> 2^{o}$ , $\rho(\theta)$ increases  with an increasing
\textbf{J}, but for small angles $\mid \theta \mid < 2^{o}$ the
opposite happens, where a minimum starts to develop with its width
increasing with an increasing applied current.} \label{fig:fig6}
\end{figure}

The angular dependence of resistivity was measured also as a
function of the applied current density at a constant temperature
and magnetic field. Fig.\ref{fig:fig6} presents the measurements
of $\rho(\theta)$ in region II for a wide range of applied current
density \textbf{J}, from 0.9 to 69.4 $kA/cm^{2}$, in a field of
0.68 T and at a temperature of 83.03 K. For angles $\mid \theta
\mid > 2^{o}$, $\rho(\theta)$ increases non-linearly with an
increasing \textbf{J}, but for small angles $\mid \theta \mid <
2^{o}$ it decreases with an increasing \textbf{J}. In the angular
region for $\mid \theta \mid < 1^{o}$, starting at small current
density ($\textbf{J} \leq 11.6 kA/cm^{2}$), the peak height
initially decreases with an increasing current, but for J larger
than $ 23.1 kA/cm^{2}$, a minimum in $\rho(\theta)$ develops. The
dependence of $\rho(\theta)$ on J is essentially the same for both
$\textbf{B} \parallel \textbf{J}$ and $\textbf{B} \perp
\textbf{J}$ orientations.

\begin{figure}[h]
\begin{center}
\def\picfilename{Fig.7.eps}
\input epsf
\epsfxsize = 210pt \epsfysize =250pt \epsfbox{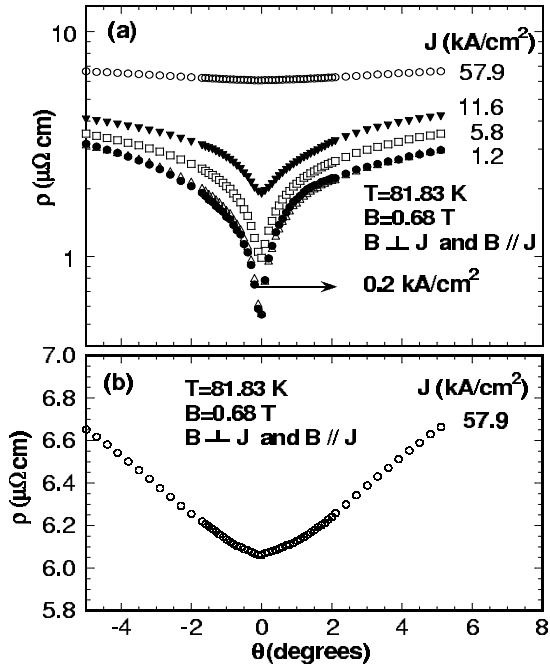}
 \end{center}
\caption{\setlength{\baselineskip}{10pt} (a) Angular dependence of
resistivity $\rho(\theta)$ measured as a function of an applied
current density \textbf{J} in a field of 0.68 T and  at a
temperature of 81.83 K (region I). The sharpness of the minimum at
$\theta=0^o$ decreases gradually with an increasing J. (b)
Expanded view of $\rho(\theta)$ measured at $J=57.9 kA/cm^2$.}
\label{fig:fig7}
\end{figure}

Fig.\ref{fig:fig7} shows $\rho(\theta)$ measured in region I for a
wide range of J in a field of 0.68 T and  at a temperature of
81.83 K. Effect of the current on the minimum is different from
that observed in region II. The minimum at $\theta=0^o$ decreases
with an increasing J. The dependence of $\rho(\theta)$ on J is
identical for both $\textbf{B} \parallel \textbf{J}$ and
$\textbf{B} \perp \textbf{J}$ orientations. Similar dependence of
$\rho(\theta)$ on J was observed over the temperature range in
region III.

\section{DISCUSSION}

The experimental data for $\rho(T,B)$ obtained for the underdoped
YBCO film are qualitatively similar to those observed before in
HTSC \cite{ref:Crusellas,ref:Mosqueira,ref:Han}. The anomalous
resistive peak is located at a temperature approximately 1.8K
lower than the onset $T_c$. The peak disappears when a magnetic
field is applied along the c-axis of the film. Also the magnitude
of the peak decreases and its position shifts to lower
temperatures with an increasing applied transport current. Our
measurements of the angular dependence of resistivity in a
magnetic field provided very valuable additional information,
which allows us to understand better the physics of the anomalous
resistivity. The measurements of $\rho(\theta)$ in a magnetic
field as a function of temperature revealed sharp minima in
resistivity at $\theta=0^o$ (\textbf{B} parallel to the ab-planes)
at temperatures below and above the resistive peak in $\rho(T)$,
and a sharp maximum at $\theta=0^o$ at the peak's temperature (see
Fig.\ref{fig:fig2}).

The data for $\rho(T)$ and $\rho(\theta)$ were used to distinguish
between different interpretations of the resistive anomaly. We
considered two-dimensional resistor model, magnetic flux motion,
and thermally activated phase-slips.

\subsection{Two-dimensional resistor model}
 In an inhomogeneous superconductor different parts of the
film or the crystal can have slightly different transition
temperatures. In this case the superconductor may be modelled as
an electrical circuit- array of different resistors. The anomalous
peak in $\rho(T)$ is produced by solving numerically, through the
standard matrix method, the electrical circuit equations. This
model was used to analyze the data, taken in a zero magnetic
field, for anomalous resistivity in $\rho_{ab}(T)$ of YBCO crystal
by Mosqueira et al \cite{ref:Mosqueira}, and in $\rho(T)$ of
Nb-based LTSC films by Vaglio et al \cite{ref:Vaglio}. The
calculated $\rho(T)$ agrees with the experimental data.

Mosqueira et al \cite{ref:Mosqueira} also attempted to explain the
suppression of the anomalous peak by a magnetic field in YBCO
crystals using the resistor model. They argued that the reduction
of the peak and its shift to low temperature is caused by the
broadening of the superconducting transition $\Delta T_c(B^*)$ in
a magnetic field $B^*$ (and not just by the shift in $T_c$). The
peak disappears when the superconducting transition broadens by
about 2K in a field of 0.3T. According to the resistive model, the
broadening required for the peak to be eliminated from $\rho(T)$
should be material-independent provided that the ratio $\Delta
\rho/\rho_B=(\rho_P -\rho_B)/\rho_B$ (where $\rho_P$ is the peak's
resistivity and $\rho_B$ is the resistivity measured in a magnetic
field \textbf{B} at the peak's temperature in the absence of the
peak) does not change. The experimental data in Refs
\cite{ref:Crusellas,ref:Mosqueira,ref:Han,ref:Santhanam} show that
this is not the case. Table \ref{tab:tab1} lists the data for
$\Delta \rho/\rho_B$, $\Delta T_c(B^*)$, and the magnetic field
$B^*$ at which the peak disappears, for different superconducting
materials.

\begin{table}
 \centering \caption{$\Delta \rho/\rho_B$, $\Delta
T_c(B^*)$, and the magnetic field $B^*$ at which the peak
disappears, for different superconductors.} \vspace{5 mm}

\begin{tabular}{c c c c c}
\hline \hline
 Material & $\Delta \rho/\rho_B$ & $B^*(T)$ & $\Delta
T_c(B^*)(K)$ & Ref \\ \hline
YBCO crystal & 0.29&0.3& 2 & \cite{ref:Mosqueira} \\
YBCO film & 0.44 & 0.08 & 0.5& this work  \\
BSCCO crystal &1.56& 0.01& $\sim 2$&\cite{ref:Han}  \\
NdCeCuO crystal &0.28& 0.1 & $\sim 4$& \cite{ref:Crusellas}\\
PrCeCuO crystal&0.16&0.7& $< 0.5$& \cite{ref:Crusellas}\\
Al wires & 0.17-0.56 & 0.001  & 0 & \cite{ref:Santhanam} \\

\hline \hline

\end{tabular}
\label{tab:tab1}
\end{table}

These data reveal that the disappearance of the peak from
$\rho(T)$ in a magnetic field is not related to the broadening
$\Delta T_c(B^*)$ of the superconducting transition. Our data
obtained on YBCO film (see Table \ref{tab:tab1}) support this
conclusion.

\subsection{Flux-motion}
\begin{figure}[h]
\def\picfilename{Fig.5f.eps}
\input epsf
\epsfxsize = 210pt \epsfysize =250pt
 \epsfbox{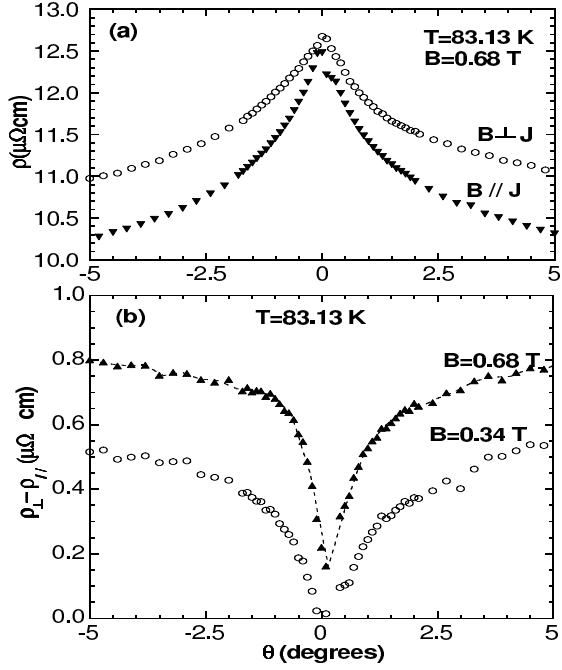}
\caption{\setlength{\baselineskip}{3mm} (a) Comparison between
$\rho(\theta)$ measured for $ \textbf{B} \perp \textbf{J}$ and $
\textbf{B} \parallel \textbf{J}$ orientations in 0.68T at 83.13K.
(b) The difference $\Delta \rho(\theta)$ between the peaks in
$\rho(\theta)$ for $ \textbf{B} \perp \textbf{J}$ and $ \textbf{B}
\parallel \textbf{J}$ orientations measured in 0.34 T and 0.68 T.}
\label{fig:fig8}
\end{figure}

In order to find the contribution of the magnetic flux motion to
the observed resistive anomaly, we performed the measurements of
$\rho(T)$ and $\rho(\theta)$ for two different orientations of the
current relative to the magnetic field i.e for $\textbf{B} \perp
\textbf{J}$ and $ \textbf{B} \parallel \textbf{J}$ orientations
(see Fig.\ref{fig:fig1}). The magnitude of the peak in $\rho(T)$
(region II in Fig.\ref{fig:fig1}) increases when the field is
parallel to the ab-planes and perpendicular to the current i.e for
$\textbf{B} \perp \textbf{J}$ (compared to the case for $
\textbf{B} \parallel \textbf{J}$). This situation corresponds to
the maximum Lorenz force acting on the flux lines along the
ab-planes. The angular dependence of $\rho$ in a magnetic field
(Fig.\ref{fig:fig2}) reveals a maximum in region II, but sharp
minima at temperatures below and above the peak (regions I and III
in Fig.\ref{fig:fig1}). A very sharp minimum in $\rho(\theta)$ at
$\theta=0^o$ ($\textbf{B}
\parallel$ ab-planes) was seen previously in a YBCO single
crystal by Kwok et al \cite{ref:Kwok} and interpreted as due to
the lock-in transition of the flux lines trapped between the
planes. For a system of weakly coupled $CuO_2$ layers one expects
a maximum resistive dissipation for $\textbf{B} \parallel $ c-axis
and a minimum for $\textbf{B} \parallel $ ab-planes. An increase
in $\rho(\theta)$ when the field is rotated from the ab-planes to
the c-axis is normally attributed to the intrinsic anisotropy of
the material. We found that the minima in $\rho(\theta)$ at
$\theta=0^o$ (regions I and III) are independent of the
orientation of the current relative to \textbf{B} (see
Fig.\ref{fig:fig4}), suggesting very strong flux lock-in mechanism
when \textbf{B} is parallel to the ab-planes. For $\theta > 0^o$
resistivity could arise via the nucleation and motion of kinks
along the vortex lines \cite{ref:Kwok}. In this case, one could
also describe the tilted vortex line as the combination of
Josephson strings aligned along the ab-planes and mobile pancakes
(vortex segments along the c-axis). If the coupling between the
pancake vortices is weak, the Lorenz force acting on these
vortices and consequently their motion, should be independent of
the direction of the transport current in the ab-planes. The
measurement of the minimum in $\rho(\theta)$ also revealed an
increase of the resistivity with an increasing transport current
in the ab-planes (Fig.\ref{fig:fig7}), which is independent of the
orientation of the current relative to \textbf{B}. This result
suggests that the motion of the pancake-vortices in the ab-planes
is responsible for the observed increase of $\rho(\theta)$ for
$\theta > 0^o$ in regions I and III. The maximum in $\rho(\theta)$
at $\theta=0^o$ at temperatures corresponding to region II in
$\rho(T)$ (Fig.\ref{fig:fig5}) depends on the orientation of the
current relative to \textbf{B}. For $\textbf{B} \perp \textbf{J}$
orientation, the maximum is higher than that measured for
$\textbf{B} \parallel \textbf{J}$ orientation. This behavior is
different from that observed in regions I and III and therefore it
provides additional argument that the peak in $\rho(T)$ can not be
explained by the 2D-resistor model alone. It also suggests that
the unknown dissipation in region II weakens flux lock-in between
the planes. Subtracting the maximum in $\rho(\theta)$ at
$\theta=0^o$ for $\textbf{B}
\parallel \textbf{J}$ from that measured for $ \textbf{B} \perp
\textbf{J}$ (see Fig.\ref{fig:fig8}) gives $\rho(\theta)$ with a
minimum similar to those observed in regions I and III, which are
caused by flux motion.

The measurement of the maxima in $\rho(\theta)$ at $\theta=0^o$ as
a function of magnetic field for $\textbf{B}
\parallel \textbf{J}$ and $\textbf{B} \perp \textbf{J}$ orientations shows
that the maxima become sharper (i.e their width decreases) with an
increasing magnetic field. For both $\textbf{B} \parallel
\textbf{J}$ and $\textbf{B} \perp \textbf{J}$ orientations, and
for $\theta>1.5^o$, the resistivity at a fixed $\theta$ decreases
with an increasing field (Fig.\ref{fig:fig5}). On the other hand,
the maximum in $\rho(\theta)$ at $\theta=0^o$ also decreases with
an increasing transport current (Fig.\ref{fig:fig6}). This
reduction in resistivity can not be explained by the flux motion.
Chaparala et al \cite{ref:Chaparala} observed a small maximum in
$\rho(\theta)$, when the magnetic field was oriented parallel to
the ab-planes in Tl (2212, 1223) and Bi(2212) crystals. The
authors did not present any data for the corresponding temperature
dependence of resistivity. The maximum in resistivity was
attributed to the formation and motion of the c-axis-oriented
vortex-antivortex segments of the flux lines parallel to the
ab-planes. They assumed that the maximum is created as a result of
the interplay between the density $n_s$ and the velocity $v_s$ of
the vortex-antivortex segments. The resistive potential difference
V is proportional to the product of these quantities. According to
the experimental observation $V \propto (n_s v_s)_{\theta=0^o}$ is
larger than $V \propto (n_s v_s)_{\theta>0^o}$. Chaparala et al
\cite{ref:Chaparala} argued that at $\theta=0^o$, in spite of the
small density of the vortices, $n_s v_s$ is large because of the
high velocity of the newly created vortex-antivortex pairs. At
$\theta>0^o$, $n_s$ is large but $v_s$ is small, so $(n_s
v_s)_{\theta=0^o} > (n_s v_s)_{\theta>0^o}$. According to this
interpretation, increasing the applied transport current should
increase the Lorenz force on these pairs and consequently increase
their velocity. This leads to an increase in the resistive
dissipation and to the growth of the maximum in $\rho(\theta)$ at
$\theta=0^o$ with an increasing current. Our data revealed a
reduction of the maximum in $\rho(\theta)$ at $\theta=0^o$ with an
increasing current (see Fig.\ref{fig:fig6}), which eliminates the
vortex-antivortex model as a possible explanation of the resistive
anomaly. The absolute values of the resistivity in the peak
observed on $\rho(T)$ curve is higher than the resistivity at the
onset $T_c$ (85 K), defined as the temperature above which the
resistivity is independent of the magnitude and direction of the
applied magnetic field (see Fig.\ref{fig:fig1}). The resistive
dissipation due to a vortex motion can reduce the critical current
density to zero, reaching the normal state resistivity, but it can
not exceed this value.

\subsection{Phase-slip model}

Discussion of the resistive-peak anomaly in the previous sections
indicates that 2D-resistor and flux motion models alone can not
fully account for the origin of this phenomenon. Regarding LTSC,
Moshchalkov et al \cite{ref:Moshchalkov} argued that the resistive
anomaly, seen in 1D Al wires, originate from thermally activated
phase slips, and the observed resistive peak at the onset $T_c$ is
the result of the phase-slip resistivity and the normal state
resistivity acting in series. Observation of the similar resistive
peak anomaly in LTSC disordered films implies that in some
disordered systems filamentary flow of the transport current could
occur through 1D constrictions (channels). We believe that this
could also happen in HTSC samples. Browning et al
\cite{ref:Browning} revealed that in spite of a large variation of
the oxygen content ($7-\delta=6.8-7.0$) measured across YBCO
crystals, they still display sharp superconducting transitions
($\sim 0.2K$), high $T_c$, and low resistivity. This implies
filamentary flow of the current in the samples. However, the
resistive peak anomaly is absent in the samples which could mean
that the filamentary flow alone is not sufficient to produce the
resistive anomalies. We conclude by analogy to the case of LTSC
disordered films that the thermally activated phase-slips  could
produce such anomaly if the filamentary flow of the current occurs
through 1D constrictions. This could happen more likely in
underdoped HTSC samples due to phase separation-induced disorder.
It should be also noted that our data show all qualitative basic
characteristics expected by the LA-MH phase slip model
\cite{ref:McCumber,ref:Langer}. According to this model phase slip
events lead to the appearance of a resistance in 1D
superconducting wires below $T_c$. During a phase slip event,
thermal fluctuations reduce the superconducting order parameter,
defined as $\psi(x)=\mid\psi(x)\mid e^{i\phi(x)}$, where $\phi(x)
$ is the phase, to zero at some point along the wire momentarily
disconnecting the phase coherence. This allows the relative phase
across the wire to slip by $2\pi$ (before $\phi(x) $ recovers its
finite value), resulting in a resistive voltage.

For a 1D thin wire with a transverse dimension $d \ll \xi$ and
$d\ll\lambda$, the LA-MH theory predicts that the appearance of a
resistance in the superconducting state is mainly determined by
thermally activated phase slips events as the system is passing
over a free energy barrier $\Delta F_{0}$ (the difference in free
energy between the normal and superconducting states) proportional
to the cross-sectional area $A$ of the wire;
\begin{equation}\label{eq:dF0}
 \Delta F_{0}=\frac{8\sqrt{2}}{3}[A\xi(T)H^{2}_{c}(T)/8\pi],
 \end{equation}
 where $H_{c}(T)$ is the thermodynamic critical field.

In the absence of the current, phase slips by $\pm2\pi$ are
equally likely, and this results in a fluctuating noise voltage
with a zero net dc component. The result of the application of a
current to the wire is to make the phase jumps more probable in
one direction than in the other. The different jump rates arise
from a difference $\delta F$ in the energy barrier for jumps in
two directions and this difference stems from the electric work
$\int IV dt$ done in the process. For a phase slip of $2\pi$, the
energy difference is $\delta F = \Delta F_{+}- \Delta
F_{-}=\frac{h}{2e}I_{s}=\phi_{0}I_{s}$, where
$\phi_{0}=\frac{h}{2e}$ is the superconducting flux quantum.
$\delta F=\phi_{0}I_s$ should be larger than the thermal energy $
k_{B}T$,
 which defines the characteristic current
$I_{1}=\frac{k_{B}T}{\phi_{0}}$, above which most phase slips go
in the driven direction and the resistance is nonlinear
\cite{ref:Tinkham}. $I_{1}$ sets a lower limit on the applied
current $I_{s}$.  The upper limit is set by the critical current
$I_{c}$ which is the mean-field critical current given by
 $I_{c}=\pi\sqrt{\frac{2}{3}}\frac{\Delta F_{0}}{\phi_{0}}$, and
\begin{equation}\label{eq:I1}
  I_{1}=\frac{k_{B}T}{\phi_{0}}<I_{s}<I_{c}
\end{equation}

The average voltage $V_{s}$ arising from the phase slip events is
determined by the number of these events in the sample
[$N(T)=L/\xi(T)$, where L is the length of the wire], a
characteristic time $\tau(T)$, Boltzmann factor $\exp(-\Delta
F(T)/k_{B}T)$, and the factor $\sinh(I_{s}\phi_{0}/2k_{B}T)$
derived from the difference $\delta F$ in the energy barrier for
the $+2\pi$ and $-2\pi$ phase jumps. $V_s$ is determined by
\begin{equation}\label{eq:Voltage}
  V_{s}=2\phi_{0}\Omega(I_{s},T) \exp{\left[-\frac{\Delta
F(T)}{k_{B}T}\right]}\sinh\left(\frac{I_{s}}{2I_{1}}\right) 
\end{equation}
where $\Delta F(T)=\Delta F_{0}(T)+(\frac{2}{3})^{1/2}I^2_s k_{B}T
/3\pi I_1 I_c$ and $\Omega(I_{s},T)$ is an attempt frequency which
can be approximated as
\begin{equation}\label{Omega}
 \Omega(I_{s},T)=\frac{N(T)}{\tau(T)}\sqrt{\frac{\Delta
F_{0}}{k_{B}T}}(1-\frac{2I_{s}}{3I_{c}})^{15/4}
\end{equation}
It is very important to emphasize the fact that the energy being
supplied during the occurrence of these phase slips at a rate of
IV, is dissipated as heat rather than converted into kinetic
energy of supercurrent, which would otherwise soon exceed the
condensation energy \cite{ref:Tinkham}.

The magnetic field dependence of the phase-slip event does not
appear explicitly in Eq.(\ref{eq:Voltage}), however its effect on
the phase-slip voltage appears through the dependence of the
critical current $I_c$ on B [$I_c \propto (1/B)$]. Phase-slip
resistivity is present over a range of the applied current $I_s$
between $I_1$ and $I_c$ according to Eq.(\ref{eq:I1}). The applied
current $I_s$ per 1D current channel should be larger than
$I_1=k_BT/\phi_0$. If $I_s$ is too close to $I_c$, the phase-slip
events are less likely to occur. Also, reducing $I_c$ while
keeping $I_s$ fixed leads to the reduction of the phase-slip
events and consequently the voltage $V_s$. For an anisotropic
superconductor, increasing the magnitude of the c-axis component
of \textbf{B} by increasing the angle $\theta$ and/or the
magnitude of \textbf{B}, reduces $I_c$.

\begin{figure}[h]
\begin{center}
\def\picfilename{Fig.10nn.eps}
\input epsf
\epsfxsize = 250pt \epsfysize =350pt
 \epsfbox{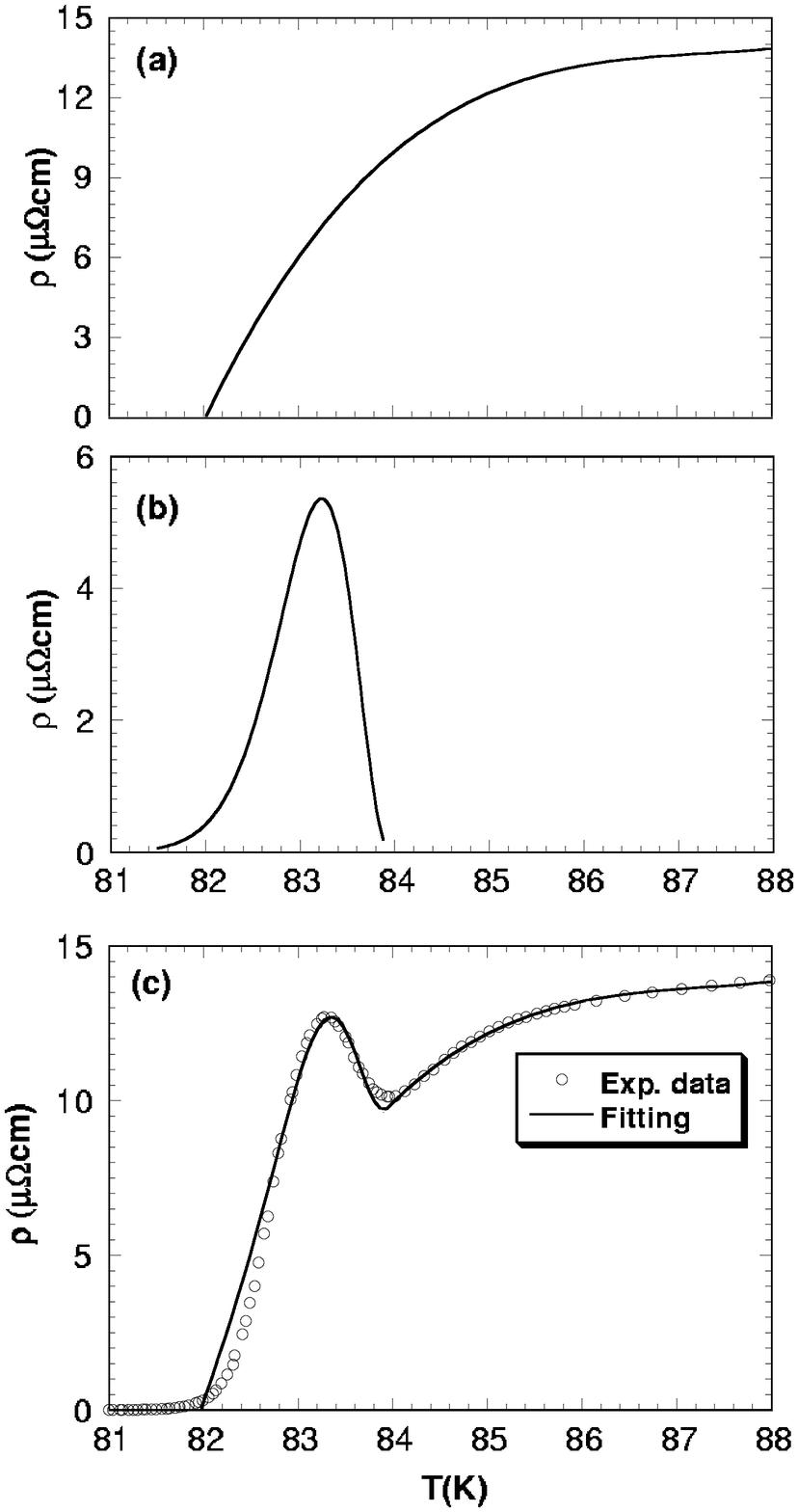}
 \end{center}
\caption{\setlength{\baselineskip}{3mm} (a) The assumed normal
state resistivity without phase-slip resistivity contribution in a
zero magnetic field; (b) The calculated phase-slip resistivity
;(c) The fitting to the experimental data obtained by
superposition of the normal state resistivity in (a) and the
calculated phase-slip resistivity in (b).} \label{fig:fig9}
\end{figure}

The angular dependence of resistivity in a magnetic field measured
over a temperature range between 82K and 84K (Fig.\ref{fig:fig2})
points out different origin of resistivity in the peak in
$\rho(T)$ (Fig.\ref{fig:fig1}), in comparison to that at
temperatures below and above the peak. Therefore $\rho(T)$ could
be treated as a superposition of the peak and the normal
resistivity near the transition, which increases almost linearly
with temperature between $T_c(R=0)\simeq 82 K$ and the onset
$T_c\simeq 85K$. We considered the possibility that the resistive
peak originates from thermally activated phase-slips and attempted
to perform numerical calculations of the phase-slip resistivity
using the modified LA-MH theory. We assumed that in an underdoped
HTSC sample the current flows through $n$ parallel superconducting
filaments of length 0.4 mm (which is the distance between the
voltage contacts). The width $w$ (in the ab-planes) and the
thickness $t$ (along the c-axis) of a filament were chosen to be
2.0 nm and 1.0 nm, respectively. These values are much smaller
than the coherence length in the ab-planes and along the c-axis at
temperatures close to $T_c$, and therefore the filaments can be
treated as 1D wires. In the system of $n$ parallel superconducting
filaments, one could expect that the phase-slip event occurring in
a filament would affect the superconducting state of the
neighboring filaments, because the coherence length at
temperatures close to $T_c$ is much larger than the spacing
between the filaments. On the other hand, in an underdoped system,
one could also expect that along each filament the superconducting
regions are interrupted by segments of normal resistance $R_N$, so
that the total resistance of the filament $R_f$ is the sum of the
phase-slip resistance and the normal state resistance acting in
series: $R_f=n_s R_s+R_N$, where $n_s$ is the number of the
superconducting segments of an average length $l_s$ and an average
phase-slip resistance $R_s$. The arguments presented above suggest
that for a system of $n$ parallel filaments, the formula for the
condensation energy $\Delta F_0$ (Eq.(\ref{eq:dF0})) for
phase-slip events in a 1D wire should be modified to reflect the
phase slip events in the whole system of $n\cdot\bar{n}_s$
segments. We assumed $\Delta F_0$ in the form:
\begin{equation}\label{eq:dF0N}
 \Delta F_{0}=\frac{8\sqrt{2}}{3}[A\xi_{ab}(T)H^{2}_{c}(T)/8\pi]\cdot n \cdot
 \bar{n}_s,
 \end{equation}
where $n$ is the total number of filaments, and $\bar{n}_s$ is the
average number of superconducting segments per filaments (the
average number of phase-slip centers per filament). The
Ginzburg-Landau expression for
$H_c(T)=H_c(0)(1-T/T_c)=H_{c2}(0)(1-T/T_c)/\sqrt{2} \kappa$, and
$\xi_{ab}(T)=\xi_{ab}(0)/(1-T/T_c)^{1/2}$ close to $T_c$ were
used. The phase-slip voltage $V_s$ was calculated using
Eq.(\ref{eq:Voltage}) and the modified attempt frequency
$\Omega(I_s,T)$. The number of the phase-slip events in a segment
was given by $N(T)=l_s/\xi_{ab}(T)\tau(T)=\gamma L/\xi_{ab}(T)
\tau(T)$, where $\gamma=l_s/L$. $\gamma$ and $\bar{n}_s$ were
treated as the fitting parameters.

The result of the calculation of $R_s=V_s/I_s$ is shown in
Fig.\ref{fig:fig9} for the following parameters : $I_s= 1 \mu A$,
${H_{c2}}_{ab}(0)=674 T$ \cite{ref:Kwok1}, $\xi_{ab}(0)=2.4 nm$
\cite{ref:Kwok1}, $\kappa=\frac{\lambda_{ab}(0)}{\xi_{ab}(0)}=58$
\cite{ref:Kwok1}, $n=4.5 \times 10^{6}$, $T_c=84.4 K$, $L=0.4
mm$,$ A=(2nm)\times (1nm)$, $\gamma= 0.067$ and $\bar{n}_s=8$.
According to Fig.\ref{fig:fig1}, the peak does not contribute to
$\rho(T)$ at temperatures above approximately 84 K, which
corresponds to $V_s=0$. The LA-MH theory does not apply at
temperatures very close to $T_c$ because of the condition for the
applied current $I_s$ which must be smaller than $I_c$
(Eq.(\ref{eq:I1})). Therefore in the calculation we used $T_c$
about 0.4 K higher. Good agreement between the experimental data
and the calculated resistivity versus temperature was obtained
(see Fig.\ref{fig:fig9}).

\begin{figure}[h]
\begin{center}
\def\picfilename{Fig.11.eps}
\input epsf
\epsfxsize = 250pt \epsfysize =300pt
\epsfbox{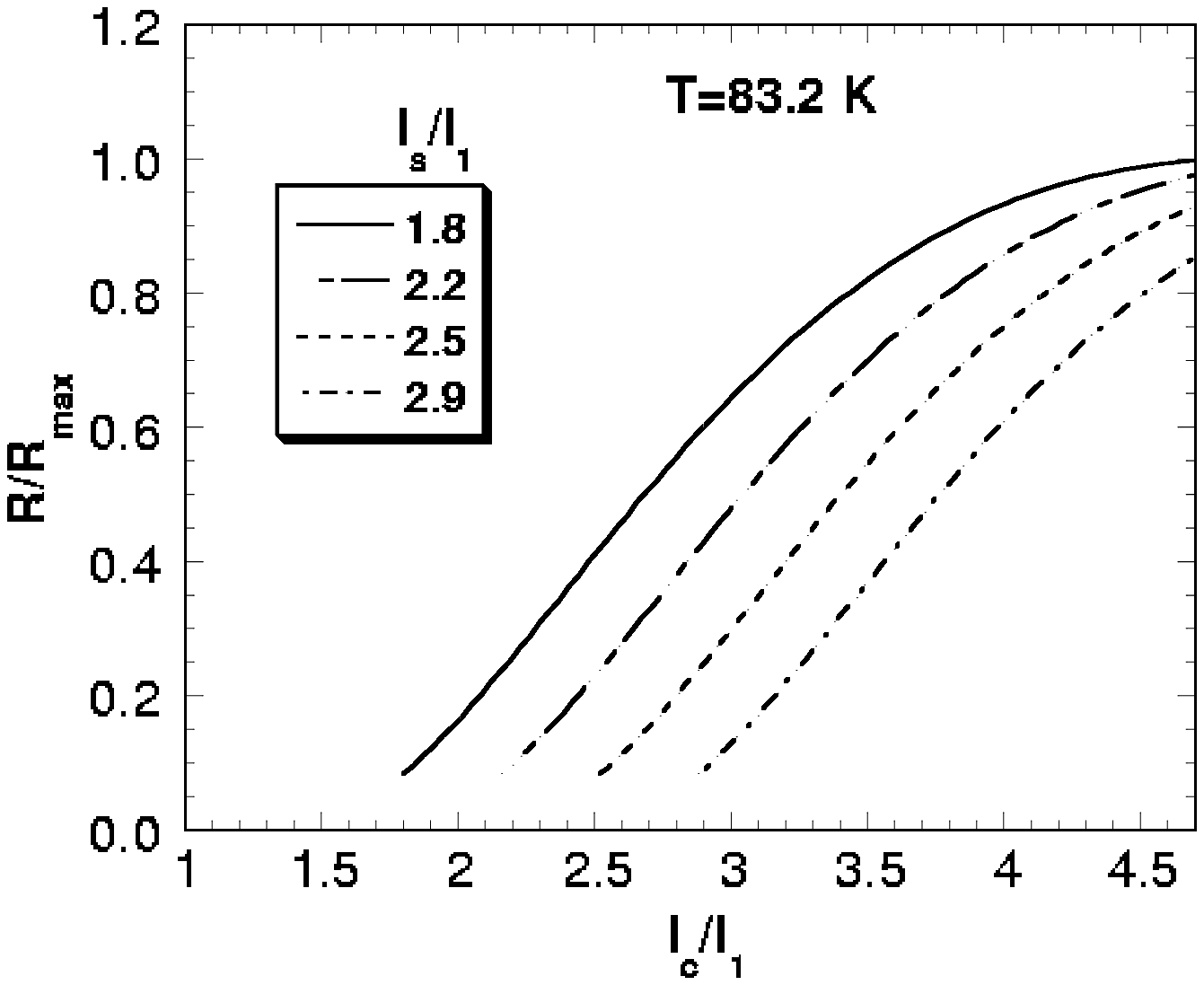}
 \end{center}
\caption{\setlength{\baselineskip}{3mm} The calculated phase-slip
resistivity as a function of the normalized critical current
($I_c/I_1$) at different applied currents $I_s$.}
\label{fig:fig10}
\end{figure}

The experimental data show that the reduction of the resistive
peak magnitude in $\rho(T)$ and the width of the peak in
$\rho(\theta)$ at $\theta=0^o$ occurs when the magnetic field
direction is rotated from the ab-planes ($\theta=0^o$) towards the
c-axis ($\theta=90^o$) (see Fig.\ref{fig:fig1}). When $B$ is
rotated from $\theta=0^o$ position, its c-axis component $B \sin
\theta$ increases and the critical current $I_c$ in the ab-planes
decreases. The resistive peaks in $\rho(T)$ and $\rho(\theta)$ at
$\theta=0^o$ also decrease in magnitude with an increasing applied
transport current $I_s$ (see Fig.\ref{fig:fig7}). We verified,
using numerical calculations that according to the LA-MH theory,
the phase-slip voltage decreases with an increasing $I_s$ and
decreasing $I_c$ in the limit of very small currents (see
Fig.\ref{fig:fig10}).

\section{SUMMARY AND CONCLUSIONS}

 We investigated the resistive peak anomaly in underdoped YBCO,
 which was observed in both the temperature dependence of
 resistivity $\rho(T)$ and the angular dependence of resistivity $\rho(\theta)$
 in an applied magnetic field $\textbf{B}$. The resistive
 peak anomaly in $\rho(T)$ decreases with an increasing
 $\textbf{B}$ (applied parallel to the c-axis) and with an increasing applied transport current $I_s$.
 On the other hand, the width of the resistive peak in
 $\rho(\theta)$ at $\theta=0^o$ decreases with an increasing
 $\textbf{B}$, and its magnitude decreases with an increasing
 $I_s$. The resistive peak anomaly in $\rho(T)$ and its dependence
 on $\textbf{B}$ and $I_s$ show striking qualitative similarities
 to those exhibited by LTSC wires, and some LTSC thin films
 and HTSC crystals. The YBCO film that we analyzed has resistivity
 much lower than YBCO crystals studied by Mosqueira et al
 \cite{ref:Mosqueira}, suggesting that the resistive peak anomaly
 is not directly related to the absolute value of the normal state
 resistivity. The anomaly can not be explained by the c-axis
 misalignments, since they would eliminate the sharp minimum in
 $\rho(\theta)$ at $\theta=0^o$ ($\textbf{B} \parallel$ ab-planes) observed in regions I and III.

 We analyzed the data in terms of three different models
 that were developed in the past to explain the resistive
 anomalies. The 2D-resistor model and the flux motion models
 are inadequate to explain fully our data including the dependence
 on a magnetic field and an applied transport current. The phase-slip
 (LA-MH) model provides the best qualitative and quantitative
 description of the observed resistive anomalies and their
 behavior as a function of temperature T, magnetic field
 \textbf{B}, the angle between \textbf{B} and the ab-planes, and
 the applied transport current $I_s$. This model can be applied
 under the assumption that the current flows through 1D
filaments. The assumption about the filamentary flow of the
current was supported by the data of Browning et al
\cite{ref:Browning}, which revealed that in spite of a very large
variation  of the oxygen content in YBCO single crystals, they
have very sharp superconducting transitions, very high $T_c$, and
low resistivity.

\section{Acknowledgements}
 \vspace{3mm} This work has been supported  by a grant
from the Natural Sciences and Engineering Research Council of
Canada (NSERC). We are grateful to M. Denhoff for supplying us
with YBCO thin films. We benefited from discussions with W-K.
Kwok, V. Vinokur and G. Crabtree.







\begin{thebibliography}{99}
\bibitem{ref:Crusellas} M. A. Crusellas, J. Fontcuberta, S. Pinol , Phys. Rev. B {\bf 46}, 14089
(1992), L. Fabrega, M. A. Crusellas,J. Fontcuberta, X. Obradors,
S. Pinol, C.J. Van der Beek, P.H. Kes, T. Grenet, and J. Beille,
Physica C {\bf 185-189}, 1913 (1991).
\bibitem{ref:Mosqueira} J. Mosqueira, A. Pomar, A. Diaz, J. A. Veira, F. Vidal
, Physica C {\bf 225}, 34 (1994).
\bibitem{ref:Han} S. H. Han, Y. Zhao, G. D. Gu, G. J. Russell, and N. Koshizuka, Adv. in
Supercond.,{\bf 8}, 109 (1996).
\bibitem{ref:Santhanam} P. Santhanam, C.C. Chi, S.J. Wind, M. J. Brady, and J. J. Bucchignano
, \emph{Phys. Rev.
Lett.} {\bf 66}, 2254 (1991).
\bibitem{ref:Moshchalkov} V. V. Moshchalkov, L. Gielen, G. Neuttiens, C. Van Haesendonck, Y. Bruynseraede
, Phys. Rev. B {\bf 49}, 15412 (1994).
\bibitem{ref:Kwong} Y.K. Kwong, K. Lin, P.J. Hakonen, M.S. Isaacson, and J. M. Parpia,
 Phys. Rev. B  {\bf 44}, 462
(1991).
\bibitem{ref:Spahn} E. Spahn and K. Keck, Solid State Commun. {\bf 78},
69 (1991).
\bibitem{ref:Vaglio} R. Vaglio, C. Attanasio, L. Maritato, and A. Ruosi, Phys. Rev. B  {\bf 47}, 15302 (1993).
\bibitem{ref:Nordstrom} A. Nordstrom and O. Rapp, Phys. Rev. B {\bf 45}, 12577 (1992).
\bibitem{ref:Langer} J.S. Langer and V. Ambegaokar, Phys. Rev.  {\bf 164}, 498
(1967);.
\bibitem{ref:McCumber} D.E. McCumber and B.I. Halperin, Phys. Rev. B  {\bf 1},
1054 (1970).



\bibitem{ref:Browning} V. M. Browning, E. F. Skelton, M. S. Osofsky, S. B. Qadri, J. Z. Hu,
 L. W. Finger, P. Caubet , Phys. Rev. B {\bf 56}, 2860 (1997).

\bibitem{ref:Andersen} N.H. Andersen, B. Lebech, and H.F. Poulsen, Physica C {\bf
172}, 31 (1990).
\bibitem{ref:Kwok} W. K. Kwok, U. Welp, V. M. Vinokur, S. Fleshler, J. Downey, G. W. Crabtree
, Phys. Rev. Lett. {\bf 67}, 390 (1991).

\bibitem{ref:Chaparala} M. Chaparala, O. H. Chung, Z. F. Ren, M. While, P. Coppens, J. H. Wang, A. P. Hope, M. J. Naughton
Phys. Rev. B {\bf 53}, 5818, (1996).
\bibitem{ref:Tinkham} M. Tinkham,\emph{Introduction to Superconductivity}, McGraw-Hill, New York,(1996).
\bibitem{ref:Kwok1} U. Welp, W. K. Kwok, G. W.Crabtree, K. G. Vandervoort, J. Z. Liu
Phys. Rev. Lett. {\bf 62}, 1908 (1989).



\end{thebibliography}
\end{document}